\documentstyle[buckow1,12pt]{article}
\newcommand{\be}{\begin{eqnarray*}}
\newcommand{\ee}{\end{eqnarray*}}
\newcommand{\bary}{\begin{array}}
\newcommand{\eary}{\end{array}}
\newcommand{\bit}{\begin{itemize}}
\newcommand{\eit}{\end{itemize}}
\newcommand{\ra}{\rightarrow}
\newcommand{\ha}{\frac{1}{2}}
\newcommand{\lear}{\leftarrow}
\newcommand{\Pls}{$\Lambda_P$ }
\newcommand{\Plv}{$10^{19}$ GeV}

\begin {document}
\begin{flushright}
{\sc DESY-98-021} \\
January, 1998
\end{flushright}
\def\titleline{The ``Ether--world'' and Elementary Particles
} \def\authors{F. Jegerlehner\footnote[1]{ \noindent 
in Proceedings of the 31st International Ahrenshoop Symposium on the
Theory of Elementary Particles, Buckow, Germany, September 1997} } 
\def\addresses{ DESY,
Platanenallee 6, D--15738 Zeuthen, Germany } 

\def\abstracttext{

We discuss a scenario of ``the path to physics at the Planck scale''
where todays theory of the interactions of elementary particles, the
so called Standard Model (SM), emerges as a low energy effective
theory describing the long distance properties of a sub--observable
medium existing at the Planck scale, which we call
``ether''. Properties of the ether can only be observable to the
extent that they are relevant to characterize the universality class
of the totality of systems which exhibit identical low energy
behavior. In such a picture the SM must be embedded into a ``Gaussian
extended SM'' (GESM), a quantum field theory (QFT) which includes the SM but
is extended in such a way that it exhibits a quasi infrared (IR) stable
fixed point in all its couplings. Some phenomenological consequences
of such a scenario are discussed.}  \makefront
\section{The Path to Physics at the Planck Scale}
The current development in the theory of elementary particles is
largely triggered by the attempt to unify gravity with the SM
interactions at the Planck scale \Pls $\sim$ \Plv. A high degree of
symmetry is required in order to cure the problems with ultraviolet
divergences.  The well known symmetry pattern is: {\bf M--T{\small
HEORY} $\sim$ STRINGS} and all that {\bf $\lear$ SUGRA $\lear$ SUSY
$\lear$ SM}, with arrows pointing towards more symmetry, provided we
neglect symmetry breakings by masses and other soft breaking terms.

Experience from condensed matter physics and a number of known facts
suggest that a completely different picture could be behind what we
observe as elementary particle interactions at low energies. It might
well be that many known features and symmetries we observe result as a
consequence of ``blindness for details'' at long distances of some
unknown kind of medium which exhibits as a fundamental cutoff the
Planck length. The symmetry pattern thus could look like: {\bf
E{\small THER} $\sim$ Planck medium $\ra$ QFT $\simeq$ GESM $\ra$
SM}. Unlike in renormalized QFT, here the relationship between bare
and renormalized parameters obtains a physical meaning. Such ideas are
quite old (\cite{Landau,Ken,HPA,tHooft} and many others) and in some
aspects are now commonly accepted among particle physicists.  Physics
at the Planck scale cannot be described by local quantum field
theory. The curvature of space-time is relevant and special relativity
is modified by gravitational effects. One expects a world which
exhibits an intrinsic cutoff corresponding to the fundamental length
$a_P \simeq 10^{-33}$ cm. But not only Poincar\'e invariance may break
down, also the laws of quantum mechanics need not hold any longer at
\Pls. The ``microscopic'' theory at distances $a_P$ is unknown, but we
know it belongs to the ``universality class'' of possible theories
which exhibit as a universal low energy effective asymptote the known
electroweak and strong interactions as well as classical gravity. Long
distance universality is a well known phenomenon from condensed matter
physics, where we know that a ferromagnet, a liquid-gas system and a
superconductor may exhibit identical long range properties (phase
diagram, critical exponents etc.). Our hypothesis could be that there
exist some kind of a ``Planck solid'', for example. We should mention
right here that there is a principal difference between a normal solid
and a ``Planck solid''; of the latter we only can observe its long
range properties, the critical or quasi--critical behavior, its true
short range properties will never be observable since we will never be
able to built a ``Planck microscope'' which would allow us to perform
experiments at \Pls. Also the observations which tell us about the
properties of the early universe will never suffice to pin down in
detail the structure at distances $a_P$. A possible ``ether-theory''
can be only a theory of universality classes, dealing with the
totality of possible systems which exhibit identical critical
behavior. It is a non--trivial task to specify possible candidate
models belonging to the universality class which manifests itself as
the SM at low energies. Here at best, we may illustrate some points,
which allow us to make plausible the viability of such a picture. The
approach discussed here is based on the experimentally well
established physics and theory of critical phenomena which teaches us
the emergence of local Euclidean renormalizable QFT as a low energy
effective structure (see e.g.\cite{Ken,Bonn}).
\section{Low energy effective theories}
The typical example we have in mind is the Landau-Ginsburg theory as
an effective ``macroscopic'' description of a real superconductor.
For a given microscopic system we may envisage the construction of the
low energy effective theory by means of a renormalization semi--group
transformation \'a la Kadanoff (block spin picture) Wilson (cutoff
renormalization group), which is a very general physical concept in
statistical physics not restricted to QFT. Let us assume that a
possible object in the universality class of interest is described by
a classical statistical system with fluctuation variables
$S_\alpha=S_{\alpha 0}+S_{\alpha 1}$, where the $S_i$ ($i=0,1$) in
momentum space have support $S_0:\; 0 \leq p \leq \Lambda /2$ and
$S_1:\; \Lambda /2 \leq p \leq \Lambda$, respectively, and
eliminating the short distance fluctuations in the partition function
yields
$${Z=\int \prod dS_\alpha e^{-H(S_\alpha)|_g}} = \int \prod dS_{\alpha
0}\prod dS_{\alpha 1} e^{-H(S_{\alpha 0}+S_{\alpha 1})|_g} = \int
\prod dS_{\alpha 0} e^{-H^{'}(S_{\alpha 0})|_{g^{'}=Rg}}$$ where $g'$
are the effective couplings of the effective theory and the effective
theory is suitable for calculating properties of the original system
at $p<\Lambda/2$. This process of lowering the cutoff by a factor of
two may be iterated in order to find the long range (low energy)
asymptote we are looking for.

Multi-pole forces arise in a natural way in such a scenario, for which
it is important to assume ``space-time'' to have some arbitrary
dimension $d \geq 4$ (see below). Since we assume all kind of fields
and excitations living at the Planck scale, there exists a long ranged
potential behaving as $\Phi \sim - \frac{1}{r^{d-2}}$ for ($r \ra
\infty$).  For weaker decay the thermodynamic limit would not exist.
Stronger decay leads to sub-leading terms at long distances.
A multi-pole expansion leads to moments of the form
$$\partial_i \Phi = (d-2)\; \frac{x_i}{r^d}\;,\;\;
\partial_j\partial_i \Phi = (d-2)\left\{ \frac{\delta_{ij}}{r^d}
-d\frac{x_ix_j}{r^{d+2}}\right\}\;,\;\;\cdots $$
which naturally mediate interactions between fluctuation variables $q$, $A_i$,
$Q_{ij}$, $\cdots$ characterized by ``energy'' forms:
$$
H = q_1q_2 \Phi \;,\;\; -q_1A_{2i}\partial_i \Phi \;,\;\; q_1Q_{2ij}
\partial_j\partial_i \Phi\;,\;\; ~~A_{1i}A_{2j} \partial_j\partial_i
\Phi\;,\;\;\cdots\;.$$ ``Charge neutrality'' for large distances
requires $q=0$.  Dipole--dipole interaction, for example, \be H =
-\sum K_{x-y,ik} A_x^i A_y^k \ee thus have a kernel \be
\tilde{K}_{ik}(q)=m^2\:\left(d\frac{q_iq_k}{q^2}
+\delta_{ik}\right)\:+c\:\left( -dq_iq_k+q^2\delta_{ik}\right)+O(q^4)
\ee and the propagators shows a from known from the massive
gauge-boson in the 't~Hooft gauge \be
\tilde{G}_{q,ij}=\left(\tilde{K}_q\right)^{-1}_{ij}=
\left(\delta_{ij}-\frac{g+bq^2}{g+bq^2+m^2+q^2}\:
\frac{q_iq_j}{q^2}\right)\frac{1}{m^2+q^2}\;\;, \ee which demonstrates
that spin 1 gauge bosons enter in a natural way.  Note that modes are
observable only if they {\it propagate}, which implies that the
leading low $q$--terms $\bar{\psi}\gamma^\mu \partial_\mu \psi$,
$\partial_\mu \phi\partial^\mu \phi$, $F_{\mu\nu}F^{\nu\mu}$, $\cdots$
determine the normalization (wave function renormalization) and this
fixes the rules for dimensional counting. Also note that by
``renormalization'' of the Bose fields and the couplings one can
always arrange the $q^2$--term in the bilinear part to be Euclidean
invariant (Liu-Stanley theorem). A similar statement should hold for
fermions.  A detailed investigation of possible low energy structures
is very elaborate. Here we mention a simplified Ansatz (assumed to ),
which was discussed as a way to derive non--Abelian gauge--theories
from ``tree unitarity'' requirements~\cite{Unitarity}. For simplicity
we assume an Euclidean invariant action at the Planck scale. Consider
only three types of particle species: scalars $\phi_a$, fermions
$\psi_\alpha$ and vector-bosons $W_{i\mu}$ with covariant
propagators. Since here we do not refer to tree unitarity but to low
energy expansion (IR power--counting) we need consider only terms
which are not manifestly irrelevant \be \bary{ccl} {\cal L} &=&
\bar{\psi}_\alpha \left\{ L^i_{\alpha \beta} P_- + R^i_{\alpha \beta}
P_+ \right\} \gamma^\mu \psi_\beta W_{\mu i}+ \ha D_{ijk} W^k_\mu
\left( W_\alpha^j\partial_\mu W^{\alpha i}- W^i_\alpha \partial_\mu
W^{\alpha j}\right)\\ &&+ \bar{\psi}_\alpha \left\{C^{+b}_{\alpha
\beta}P_+ + C^{-b}_{\alpha \beta} P_-\right\} \psi_\beta \phi^b+\ha
K_{ij}^b W_\mu^i W_\mu^j \phi_b \\ &&+ \ha
T^i_{ba}W^i_\mu\:\left(\phi_a \partial_\mu \phi_b-\phi_b \partial_\mu
\phi_a \right) +\frac{1}{4}M^{ij}_{ab} W_{\mu i}W_{\mu j}\phi_a \phi_b
\eary \ee with arbitrary interaction matrices of the fields.  The
extraction of the leading low energy asymptote is equivalent to the
requirement of renormalizability of $S$--matrix elements, and this has
been shown to necessarily be a non--Abelian gauge theory which must
have undergone a Higgs mechanism if the gauge bosons are not strictly
massless.  Since terms of order $O(E/\Lambda_P)$ are automatically
suppressed in the low energy regime only a renormalizable effective
field theory can survive as a tail, the possible renormalizable
theories on the other hand are known and are easy to classify. Thus
gauge symmetries and in particular the non--Abelian ones appear as a
conspiracy of different modes ``self--arranged'' in such a way the
$O((E/\Lambda_P)^n)$--terms ($n>0$) are absent.  Also
anomaly--cancelation and the related quark--lepton duality (family
structure) are easily understood and natural in such a context. To an
accuracy of $E/\Lambda_P=10^{-3}$ we are thus dealing with a
renormalizable local QFT of the ``spontaneously broken gauge theory''
(SBGT) type at a scale $10^{16}$ GeV. As we shall argue below this
cannot be just the SM.

The fact that there are only a few possible forms for the low energy
effective theories is particularly attractive and tells us that
symmetries and particular mathematical properties may be interpreted
to emerge as low energy patterns.

The low energy expansion in terms of field monomials makes sense only
in the vicinity of a second (or higher) order phase transition point
where the system exhibits long range correlations and is described in
the long range limit by an effective conformal quantum field theory
characterized by an infrared stable fixed point. It is well known that
for dimensions $d>4$ such effective theories turn out to be Gaussian
(free field) theories. Non--trivial theories with a stable ground
state are possible only for dimensions $d\leq 4$. At the boarder case
$d=4$ non-trivial long range interactions set in and we expect
effective couplings to be weak and therefore perturbation theory to
work\cite{Ken}. The quasi--triviality but non-triviality in this
scenario is due to the fact that a huge but finite cutoff, namely
$\Lambda_P$, exist in the underlying physics. Such a scenario
explains why elementary particle interactions, up to scales explored so
far, are described by renormalizable quantum field theory and why we
can do perturbation theory. Note that space-time ``compactifies
itself'' by the decoupling of the $n=d-4$ extra dimensions.

In $d=4$ space-time dimensions there exist infinitely many infrared
``irrelevant'' (non--renormalizable) operators of dimension $>$ 4
(scaling like $(E/\Lambda_P)^n$ with $n\geq 1$); but there exist only
relatively few infrared ``marginal'' (strictly renormalizable)
dimension 4 operators (scaling like $\ln^n (E/\Lambda_P)$ with $n\geq
1$) and even fewer infrared ``relevant'' (super--renormalizable)
operators of dimension $<$ 4 (scaling like $(\Lambda_P/E)^n$ with
$n\geq 1$). The dimension $\leq$ 4 operators characterize a
renormalizable QFT. The relevant operators must be tuned for
criticality in order that the low energy expansion makes sense. This
fine tuning is of course the main obstacle for such a vision to be
convincing. Unlike in a condensed matter physics laboratory we cannot
tune by hand the temperature and the external fields for
criticality. However, since we have good reasons to assume that there
exist a dense variety of fluctuations and modes it is conceivable that
at long distances we just see those modes which ``conspire'' precisely
in such a way as to allow for long distance fluctuations. This
conspiracy is nothing but the ``symmetry patterns'' which emerge at
long distances.  Other existing modes just are frozen at short
distances and are not observable. The ``critical dimension'' $d=4$ is
crucial for the scenario to work because weakly interacting large
scale fluctuations are expected to govern the quasi critical
region. As we mentioned above, gauge symmetries are particularly easy
to understand within this context. The gauge groups expected in such a
scenario of course are the ones which follow by conspiracy of
``particles'' in singlets, doublets, triplets, etc. exactly as we
observe them in the real world. Thus, while a $U(1) \otimes SU(2)
\otimes SU(3) \otimes \cdots $ pattern looks to emerge in a natural
way, we never would expect higher dimensional multiplets to show up if
the particular symmetry would not be there already a the Planck
scale. Also the repetition of fermion family patterns, known in the
SM, looks to be a rather natural possibility in our approach.
\section{Natural properties at low energies}
Above we outlined that known empirical facts about structural
properties of elementary particle theory find a natural explanation in
low energy effective theories. Usually quantum mechanics and special
relativity, four dimensionality and renormalizability are independent
inputs. Detailed investigations confirm that all these properties may
be understood as consequences of the existence of an ``ether'' in the
appropriate universality class. At long distances we observe: 1.)
Local quantum field theory; note that the equivalence of Euclidean QFT
and Minkowski QFT is a general property of any renormalizable QFT
(Osterwalder--Schrader theorem). The analyticity properties allow for
the necessary Wick rotation. In its Minkowski version QFT incorporates
quantum mechanics and special relativity, which thus show up as low
energy structures more or less automatically\cite{Bonn}.  2.)
Space--time dimension $d=4=3+1$.  3.)  Interactions are renormalizable
and thus described by a Lagrangian which includes low dimensional
monomials of fields only.  4.) Weak coupling and perturbative nature
of elementary particle interactions; this is natural only if we
\underline{require} the low energy effective QFT in $d=4$ to exhibit a
trivial (Gaussian) IR fixed point, which is the natural candidate for
the low energy effective theory to stabilize. Since in addition we
have to \underline{require} the SM to part of it we call it ``Gaussian
extended SM'' (GESM). It is weakly interacting at low energies due to
the existence of the large but finite cutoff. 5.) Local gauge
symmetries with small gauge groups; they provide the dynamical
principle which fixes the interactions of the SM and of Einstein's
theory of gravity (equivalence principles).  6.) The existence of a
large finite physical cutoff implies that the relationship between
bare and renormalized quantities are physical.

Basis of all this is the equivalence of statistical mechanics near
criticality and quantum field theory. There are only a few low range
theories possible (conformal quantum field theories characterized by a
few properties like global symmetries, dimension etc.)  Also the
equivalence of the path integral quantization and the canonical
quantization is more than an accident within this context.

Further consequences are briefly discussed in the following: i)~Since
we need a quasi Gaussian IR fixed point, asymptotic freedom as seen in
the SM must be lost at higher energies; this requires $N\geq$ 9
families to exist. The asymptotic freedom of QCD and in the $SU(2)_L$
coupling are a consequence of the decoupling of the heavier fermions.
ii)~Since the relationship between bare and renormalized parameters
must be physical, positivity of counter terms etc. must be required,
which has direct consequences for vacuum stability and positivity of
both the bare and the renormalized Higgs potential, for example.  
iii)~QFT properties are expected to be violated once
$E/\Lambda_P,\;\;(E/\Lambda_P)^2, \cdots$ terms come into play; at
$E\simeq 10^{16}$ GeV one might expect 0.1\% effects. This might be
important to remember in the attempts to solve the puzzle of
baryogenesis, for example.

The simplest GESM may be obtained by adding more (heavier) fermion
families to the SM. For a SM with $N$ families the one--loop counter
terms for the $U(1)_Y$, $SU(2)_L$ and $SU(3)_c$ couplings read 
$$\delta g/g=c_g\:\ln (\Lambda^2/\mu^2) $$ with
\be \begin{array}{lcl}
c_{g'}&=&\frac{{g'}^2}{24\pi^2}\:(5/3\:N+1/8),\\
c_g&=&\frac{g^2}{24\pi^2}\:(N-11/2+1/8),\\
c_{g_s}&=&\frac{g_s^2}{16\pi^2}\:(4/3 \:N-11),
\end{array} \ee
respectively. The
corresponding $\beta$--functions must all be positive (IR fixed point
condition) in the weak coupling limit. For the Abelian coupling we
have $c_{g'}>0$ in any case. For the non-Abelian couplings $c_{g}>0$
provided $N\geq6$ and $c_{g_s}>0$ provided $N\geq9$, i.e. they must be
matter dominated. Note that for the unbroken $U(1)_{\rm em}$
$$c_e=\sin^2\Theta_W c_g+\cos^2\Theta_W
c_{g'}=\frac{e^2}{24\pi^2}\:(8/3\:N-11/2 +1/4)>0$$ for $N\geq2$.  In
our scheme there is a prediction for $${\tan^2
\Theta_W}_{\mathrm{eff}}= ({g'}^2/g^2)_{\mathrm{eff}}= \frac{{g'}^2
(1+2 \delta g'/g')}{g^2 (1+2 \delta g/g)} \sim
\frac{24N-129}{40N+3}.$$ It is positive only provided $N\geq6$ and we
obtain $\sin^2 \Theta_W$=0.05814,0.16321,0.19333,0.23356 and 0.37500
for $N$=6,8,9,11 and $\infty$. Note that more realistic estimates must
include appropriate threshold/decoupling effects. Of course the
existence of additional fermion families is possible, although
additional light neutrinos are excluded as we know.

We finally have to worry about the quadratic divergences i.e. the
tuning of the relevant parameters for criticality\cite{Quadiv}. In the SM,
utilizing dimensional regularization, the quadratic divergences show
up as poles at $d=2$ and this solely concerns the Higgs mass counter
term \cite{FJ81} 
\be \begin{array}{ccl} 
\delta m_H^2=\frac{1}{16\pi^2v^2} \{
A_0(m_H)\:3m_H^2 +A_0(M_Z)\:(m_H^2+6M_Z^2) +A_0(M_W)\:(2m_H^2+12
M_W^2) \\
+\sum_{f_s}A_0(m_f)\:(-8m_f^2)+\cdots \}
\end{array} \ee 
where
$$A_0(m)=\Lambda^2 (m^2/\mu^2)^{(d/2-1)} \left( 4\pi \right)^{-d/2}
\Gamma (1-d/2)$$ and thus $$\delta m_H^2 \sim 6 (\Lambda/v)^2
(m_H^2+M_Z^2+2M_W^2-4m_f^2 )$$ and the IR fixed point condition
requires\cite{Veltman} $$m_H \simeq (4(m_t^2+m_b^2) -M_Z^2-2M_W^2)^{1/2}\sim 318~~
\mathrm{GeV}.$$  This lowest non--trivial order
consideration seems to predict a reasonable value of the Higgs
mass. Since we are in a perturbative regime higher order perturbative
corrections modify the precise value of the prediction but they cannot
affect the existence of a solution which is not ruled out by
experiment. Actually, current precision measurements very strongly
suggest that the Higgs coupling is fairly weak (SM fits favor $m_H <
420$ GeV at 95 \% C.L. and thus $\lambda =m_H^2/(2v^2) < 1.5$ which
leads to an expansion parameter about $\alpha_\lambda
=\lambda^2/(4\pi)\sim 0.18$). In fact the symmetry which constrains
the scalar mass here is dilatation invariance. Of
course, like in SUSY theories, the cancelations of the contributions
is only possible between fermionic and bosonic degrees of
freedom. This prediction should not be taken too serious. First of all
it does not include the effects from the extra heavy fermion families
which must exist in this scheme. More serious is the expectation that
such a result cannot be universal, it is expected to depend on the
actual structure of the ``bare'' theory, which is unknown. On the
other hand renormalizable SBGT is in effect up to energies of about 
$10^{16}$ GeV and below that standard gauge invariance and RG
arguments apply.

At this point we should remember 't Hooft's naturalness argument:
``Small'' masses are natural only if setting them to zero increases
the symmetry of the system\cite{tHooft}.  Indeed a light particle
spectrum (IR relevant terms) must be the result of a ``conspiracy''
i.e. modes conspire to form approximately multiplets of some symmetry
which protects the masses from large renormalizations: light fermions
require approximate chiral symmetry, light vector bosons require
approximate local gauge symmetry, light scalars require approximate
super symmetry or approximate dilatation symmetry. Remember that
dilatation invariance implies conformal invariance.

The view developed in the previous sections has to be worked out in
more details in many respects. There are many open problems, for
example, concerning the origin of fermions. I think this is a
promising framework which should be considered seriously. One big
advantage is that it has non--trivial phenomenological consequences
which are testable in the not too far future.


\begin{thebibliography}{77}

\bibitem{Landau}
L. D. Landau, in ``Niels Bohr and the Development of Modern Physics'',
ed. W. Pauli, Mc Graw-Hill, New York, 1955, p.52

\bibitem{Ken}
K.G. Wilson, {\it Phys. Rev.} {\bf B4} (1971) 2174, ibid. 3184\\
K.G. Wilson, M.E. Fisher, {\it Phys. Rev. Lett.} {\bf 28} (1972) 240\\
K.G. Wilson, J. Kogut, {\it Phys. Rept.} {\bf 12} (1974) p. 75\\
S.D. Glazek, K.G. Wilson, {\it Phys. Rev.} {\bf D49} (1994) 4214


\bibitem{HPA}
F.~Jegerlehner, {\it Helv. Phys. Acta} {\bf 51} (1978) 783

\bibitem{tHooft}
G.~'t Hooft, in ``Recent Developments in Gauge Theories'', 
G.~'t Hooft et al. (eds.), Plenum Press, New York, 1980, p. 135

\bibitem{Bonn}
F. Jegerlehner, in
``Trends in Elementary Particle Theory'', eds. H. Rollnik, K. Dietz,
Springer, Berlin, 1975, p.114\\
J.~Kogut, {\it Rev. Mod. Phys.}{\bf 51} (1979) 659

\bibitem{Unitarity} C.H.~Llewellyn Smith, {\it Phys. Lett.} {\bf B46} (1973) 233; 
J.S.~Bell, {\it Nucl. Phys.} {\bf 60} (1973) 427;
J.M.~Cornwall, D.N.~Levin, G.~Tiktopoulos, {\it Phys. Rev. Lett.} {\bf
30} (1973) 1268, 31(E)
(1973) 572, {\it Phys. Rev.} {\bf D10} (1974) 1145.

\bibitem{FJ81} J.~Fleischer, F.~Jegerlehner,
{\it Phys. Rev.} {\bf D23} (1981) 2001 

\bibitem{Quadiv} I.~Jack, D.R.T.~Jones, {\it Phys. Lett.} {\bf B234}
(1990) 321, {\it Nucl. Phys.} {\bf B342}
(1990) 127, M.S.~Al-Sarhi, I.~Jack, D.R.T.~Jones, {\it Nucl. Phys.} {\bf B345}
(1990) 431, {\it Z. Phys.} {\bf C55} (1992) 283

\bibitem{Veltman} M.~Veltman,
{\it Acta Phys. Polonica} {\bf B12} (1981) 437, see also: Z.Y.~Fang et al., Preprint UCL-IPT-96-22 (1996), {\bf hep-ph/9612430 v2} 

\end{thebibliography}
\end{document}